\documentclass[letterpaper]{article} 
\usepackage[submission]{aaai25}  
\usepackage{times}  
\usepackage{helvet}  
\usepackage{courier}  
\usepackage[hyphens]{url}  
\usepackage{graphicx} 
\usepackage{natbib}  
\usepackage{caption} 
\usepackage{algorithm}
\usepackage{algorithmic}
\usepackage{color}
\usepackage{caption}
\usepackage{tabularx}
\usepackage{booktabs}
\usepackage{subcaption}
\usepackage{xcolor}

\usepackage{newfloat}
\usepackage{listings}
\usepackage{amsmath}
\usepackage{dsfont}
\usepackage{multirow}

\DeclareCaptionStyle{ruled}{labelfont=normalfont,labelsep=colon,strut=off} 
\floatstyle{ruled}
\newfloat{listing}{tb}{lst}{}
\floatname{listing}{Listing}
\title{Less is More: Unseen Domain Fake News Detection via Causal Propagation Substructures}

\author {
    Shuzhi Gong\textsuperscript{\rm 1},
    Richard O. Sinnott\textsuperscript{\rm 1},
    Jianzhong Qi\textsuperscript{\rm 1},
    Cecile Paris\textsuperscript{\rm 2}
}
\affiliations {
    \textsuperscript{\rm 1}The University of Melbourne\\
    \textsuperscript{\rm 2}Data61, CSIRO\\
    \{shuzhi, rsinnott, jianzhong.qi\}@unimelb.edu.au, Cecile.Paris@data61.csiro.au
}

\newcommand\method{\texttt{CSDA}}

\begin{document}

\maketitle

\begin{abstract}

The spread of fake news on social media poses significant threats to individuals and society. Text-based and graph-based models have been employed for fake news detection by analyzing news content and propagation networks, showing promising results in specific scenarios. However, these data-driven models heavily rely on pre-existing \emph{in-distribution} data for training, limiting their performance when confronted with fake news from emerging or previously unseen domains, known as \emph{out-of-distribution} (OOD) data. Tackling OOD fake news is a challenging yet critical task. In this paper, we introduce the \underline{\textbf{C}}ausal \underline{\textbf{S}}ubgraph-oriented \underline{\textbf{D}}omain \underline{\textbf{A}}daptive Fake News Detection (\method) model, designed to enhance zero-shot fake news detection by extracting causal substructures from propagation graphs using in-distribution data and generalizing this approach to OOD data. The model employs a graph neural network-based mask generation process to identify dominant nodes and edges within the propagation graph, using these substructures for fake news prediction. Additionally, \method's performance is further improved through contrastive learning in few-shot scenarios, where a limited amount of OOD data is available for training. Extensive experiments on public social media datasets demonstrate that \method~ effectively handles OOD fake news detection, achieving a 7\%$\sim$16\% accuracy improvement over other state-of-the-art models.

\end{abstract}

%

\section{Introduction}
The popularity of social media has enabled rapid  news dissemination, for both true and fake news. Given the potential impact of fake news, robust fake news detection methods are needed to debunk such news in a timely manner. In real-world scenarios, out-of-distribution news from unseen domains emerges over time. This brings substantial challenges to fake news detection models.


Graph-based fake news detection methods using graph neural networks (GNN) have garnered much attention recently for modelling  news 
propagation patterns~\cite{gong2023survey}. Despite their success, existing GNN-based methods are generally built on the assumption that both training and testing data are independently sampled from an identical data distribution (i.i.d.), which often does not hold true nor reflect the real challenges of fake news detection~\cite{li2022graphde}. Emerging and hitherto unseen fake news and their associated propagation graphs can and do appear. From an empirical perspective, these methods focus on minimising the average training error and incorporating correlations within the training data (which is considered to be \emph{in-distribution}) to improve fake news detection accuracy~\cite{liu2021towards-ood-generalisation-survey}.
However, real-world graph-based fake news data is often mixed with biased domain-specific information in the training data. The detection model may thus learn these domain-specific biases resulting in misclassification of cross-domain news items~\cite{li2022graphde}.



To detect fake news across different domains (e.g., sports and politics), some early studies~\cite{ma2018rumor,bian2020rumor} focus on capturing content-independent propagation patterns. However, it has been shown~\cite{min2022divide} that not only the news contents but also the propagation patterns can vary across different news domains. 
More recent approaches~\cite{li-acmmm-2023improving,lin-naacl-2022-aclr} collect and manually label a small dataset from emerging news domains. They utilise domain adaptation methods to adapt the trained models to the emerging domains in a few-shot manner. However, these approaches require labelled data from emerging domains which is not always available and could be expensive and time-consuming.

\begin{figure}[ht]
\centering
\includegraphics[width=0.49\textwidth]{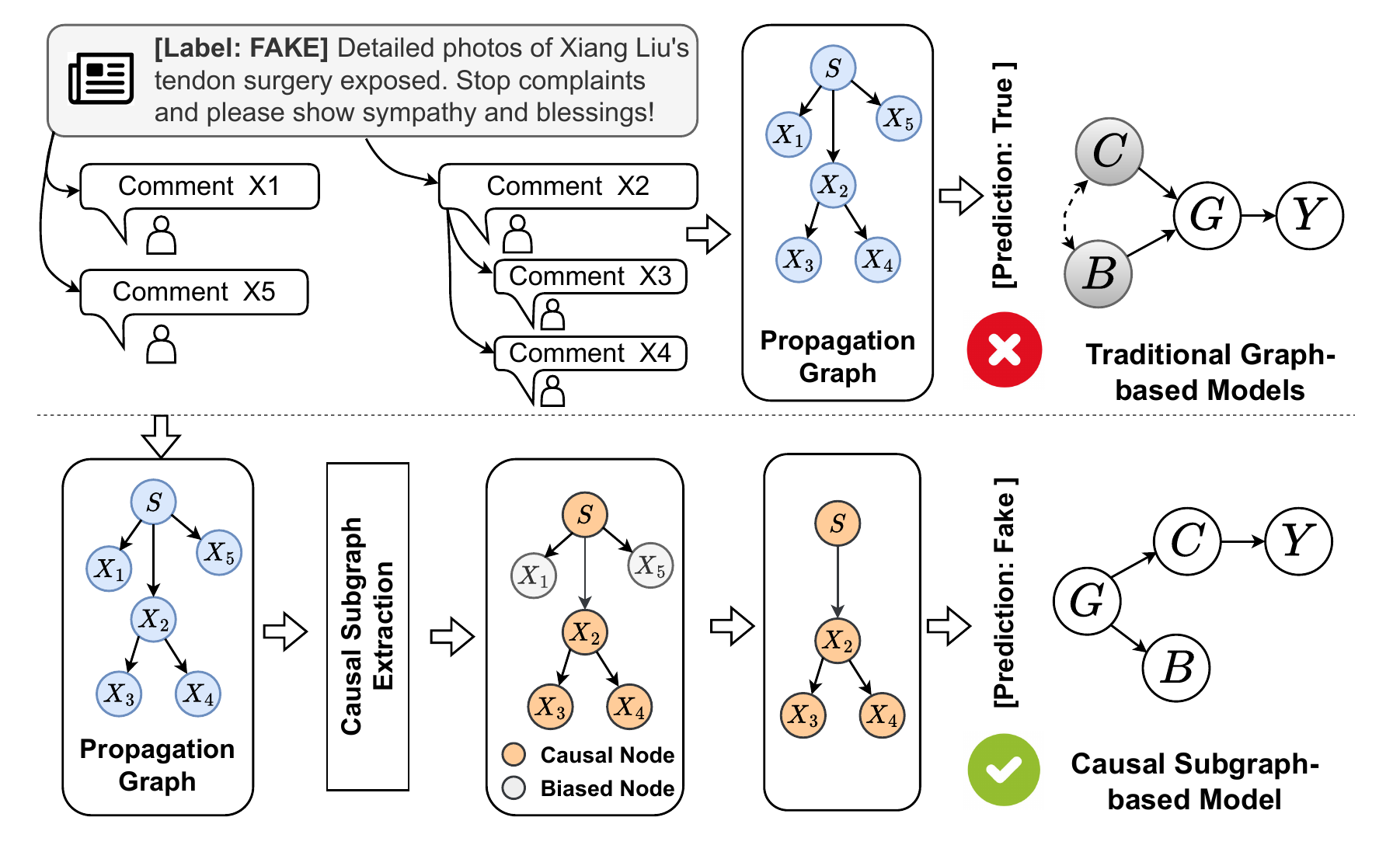}
\caption{Illustration of the causal subgraphs and the Structure Causal Models (SCMs). In the SCMs, the grey and white variables represent unobserved and observed variables. Further explanations on SCMs are given in Preliminaries.}
\label{Fig.SCM}
\end{figure}

To address the limitations above, we focus on extracting causal subgraphs from news propagation graphs to eliminate potential domain biases. The patterns of such subgraphs are learnt for fake news detection in emerging domains. News from an emerging domain is considered as the \emph{out-of-distribution} (OOD) data, and we generalise our model trained on in-distribution data to OOD data by capturing causal subgraphs in an unsupervised manner. From a causal analysis perspective, each propagation graph is composed of causal subgraph and biased subgraph which are initially entangled. Our intuition is that not all nodes in the propagation graph of a given news item are helpful for fake news detection. Instead, only some causal subgraphs of the propagation graph carry critical clues that can be used to identify fake news, as illustrated in Fig.~\ref{Fig.SCM} with an example. If we can identify and capture such causal subgraphs, we can improve fake news detection accuracy and subsequently improve the way we generalise the model to OOD data.

Based on this intuition, a cross-domain model -- the \underline{\textbf{C}}ausal \underline{\textbf{S}}ubgraph Oriented \underline{\textbf{D}}omain \underline{\textbf{A}}daptive Fake News Detection (\method) model, is proposed. This model extracts subgraphs from propagation graphs and performs detection based on the subgraphs.  In \texttt{CSDA}, a binary mask is learned for each node and each edge of the propagation graph of a news item to classify them into \emph{causal} or \emph{biased} elements. For the subgraph formed by each type of element, a graph encoder and a multilayer perceptron (MLP) classifier together encode the subgraphs and classify the news item according to the subgraph embeddings. In the training process, we utilise a data augmentation strategy by concatenating the causal subgraph embedding and the permuted biased subgraph embedding. We then train \method\ with both embeddings to enhance the effectiveness of causal subgraph learning. In the testing process, only the causal branch of the \texttt{CSDA} model is utilised to predict news veracity.

Following recent works~\cite{li-acmmm-2023improving,lin-naacl-2022-aclr}, we also consider a scenario where limited OOD data becomes available through manual labelling. In this scenario, \method's performance on OOD data is further enhanced with a supervised contrastive learning-based approach and achieves state-of-the-art (SOTA) classification accuracy.

In summary, our contributions include:
\begin{itemize}
    \item We propose a zero-shot cross-domain fake news detection model named \method\ based on extracting causal subgraphs related to news propagation patterns.
    \item We further explore a few-shot scenario in cross-domain fake news detection where a small number of OOD examples are available, and we utilise contrastive learning to enhance \method's cross-domain performance. 
    \item Extensive experiments are conducted on four real datasets. The results confirm the effectiveness of \method\ in cross-domain fake news detection, outperforming SOTA models by $7.69\sim16.00\%$ in terms of accuracy.
\end{itemize}

\section{Related Work}
\subsection{Traditional Fake News Detection Methods}
Traditional fake news detection methods can be divided into \emph{content-based},  \emph{social context-based} and \emph{environment-based}.

Content-based methods learn content or style features from the text or multi-media content of news~\cite{feng2012syntactic}. They may also leverage external knowledge for fact checking~\cite{samarinas2021improving}. Social context-based methods detect fake news through user features \cite{shu2019beyond} and users' roles in news propagation. Propagation analysis is a trending topic in social-context based methods, with models being developed for sequence modelling~\cite{ma2016detecting,khoo2020interpretable} and graph modelling~\cite{bian2020rumor}. Temporal propagation features are often exploited. For example,~\citet{choi2021dynamic} encode the propagation process as graph snapshots;~\citet{song2021temporally} utilise a temporal graph network (TGN)~\cite{xu2020inductive} to encode the propagation graph, whilst~\citet{naumzik2022detecting} and~\citet{gong2023fake} classify fake news by  different self-exciting patterns. Environment-based methods~\cite{nguyen2020fang} consider associations across multiple news to extract context for fake news detection.

\subsection{Cross-Domain Fake News Detection}

Cross-domain fake news detection aims to train a detection model in one domain (the \emph{source domain}) and apply the model to a different domain (the \emph{target domain}). To achieve cross-domain detection, existing works can be largely categorised into \emph{sample-level} and \emph{feature-level} methods. 

Sample-level methods identify domain-invariant data samples in the training set and assign larger weights to those samples~\cite{silva2021embracing,yue-cikm-2022contrastive}. Studies in this category~\cite{yue-cikm-2022contrastive,ran-aaai-2023-unsupervised} leverage clustering algorithms to augment target domain training samples and then train the models together with both source and target domain data, thereby improving the model performance on the target domain data. Feature-level methods focus on weighting or extracting domain-independent features. For example,~\citet{mosallanezhad2022domain} utilise reinforcement learning to select domain-invariant attributes from the news features. Inspired by the domain-adaptive neural networks~\cite{ganin-icml-2015-dann-unsupervised}, studies~\cite{min2022divide,li-acmmm-2023improving} train an additional domain discriminator adversarially by attempting to generate news embeddings that cannot be recognised by the domain discriminator. In this case, the generated news embeddings are considered to be domain-invariant. In this paper, we utilise more information by extracting causal propagation substructures. 

\section{Preliminaries}

Cross-domain fake news detection aims to transfer a model trained on a labelled (in-distribution) dataset to an OOD dataset that is unlabelled or with a few labelled samples. 

Given a set of news items $D_{in} = \{(\mathcal{G}_{k}^{in} , y_{k}^{in})\}$ ($k \in [1, n_{in}]$) that comes from some latent distribution $\mathcal{P}$, we aim to train a model to detect fake news in another dataset $D_{out}=\{(\mathcal{G}_{k}^{out})\}$ ($k \in [1,n_{out}]$) that contains data from an unknown distribution $\mathcal{P^{'}}$ different from $\mathcal{P}$. Here, we refer to data from $\mathcal{P}$ as in-distribution data and those from $\mathcal{P^{'}}$ as OOD data. $D_{in}$ is the in-distribution data and $D_{out}$ is the out-of-distribution data, while 
$n_{in}$ and $n_{out}$ refer to the number of news items in $D_{in}$ and $D_{out}$, respectively. 
Our goal is to train a classifier $f$ using the  training set $D_{in}$ to determine whether news items in another non-over-lapping set $D_{out}$ contain fake news. We assume that both $D_{in}$ and $D_{out}$ share the same label space.

\noindent \textbf{Causal Analysis:} As shown in Fig.~\ref{Fig.SCM}, from a causal view, the variables $C$, $B$, $G$, $Y$ represent the casual subgraph, the biased subgraph, the observed propagation graph, and the news label. Each link denotes a causal relationship~\cite{DisC-fan2022debiasing}. In the traditional graph-based models, the propagation graphs are encoded directly therefore the spurious correlation between $C$ and $B$ is ignored and fused into the graph embedding, leading to inaccurate prediction. In our causal subgraph extraction, the causal subgraphs and biased subgraphs are disentangled, and the prediction can be improved by referring solely to the causal information.

\noindent\textbf{Data preparation:} 
For each news item from both $D_{in}$ and $D_{out}$, its propagation graph $\mathcal{G}_{k} =\langle \mathbf{X}_{k}, \mathbf{A}_{k} \rangle$ is extracted and modelled as a directed acyclic graph. The node set $\mathbf{X}_{k} = \{x_1, x_2, \ldots,x_{|\mathbf{X}_{k}|}\}$ contains all posts including the source news post and all associated comments/reposts which provide supportive information about the post veracity. Each post's embedding is initialised using a pre-trained BERT model~\cite{devlin2018bert} to compute the text embeddings.

The adjacency matrix $\mathbf{A}_{k} =\{\alpha_{mn}, m, n \in [1, |C_{k}|]\}$ is the set of propagation behaviours where an edge exists (i.e., $\alpha_{mn}=1$) between node $m$ and node $n$ if there is a reply/repost relationship.

\section{Proposed Model}

\begin{figure*}[ht]
\centering
\includegraphics[width=\textwidth]{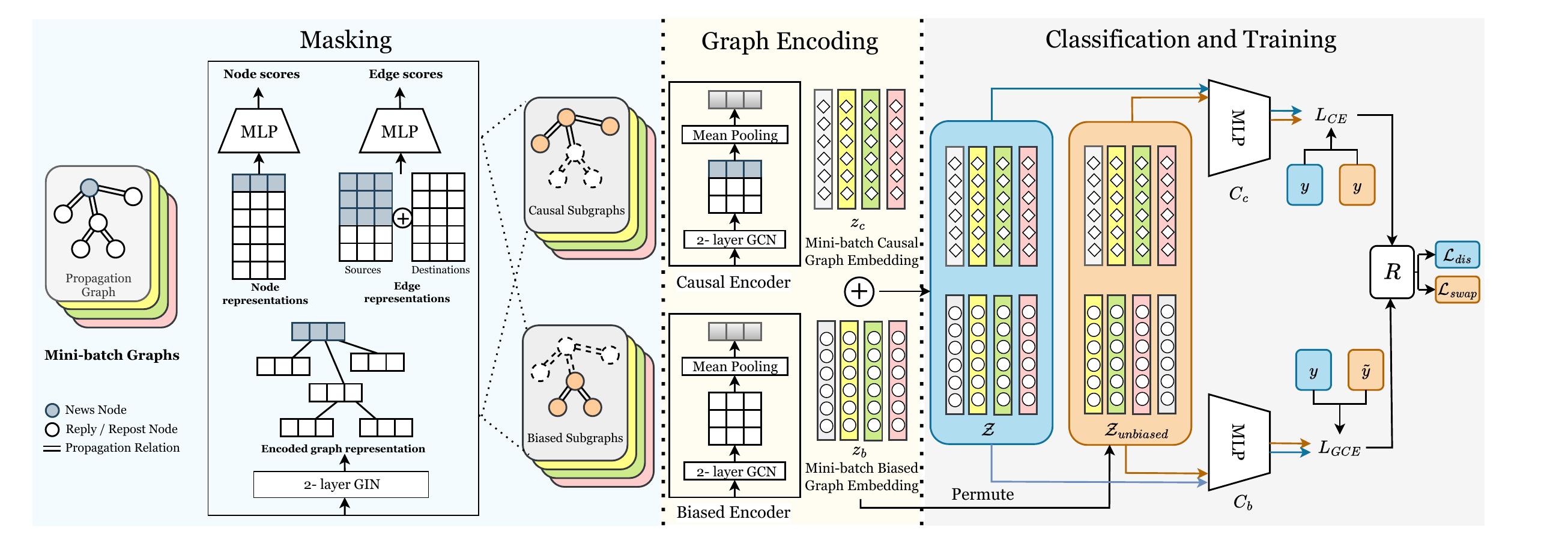}
\caption{Architecture of \method, which is trained with batches of news propagation graphs. A mini-batch of propagation graphs are masked by the Mask Generator and divided into causal subgraphs and biased subgraphs. Then, the two batches of subgraphs are encoded by two independent graph encoders into causal and biased embeddings. Afterwards, two MLP-based prediction modules that focus on the causal and the biased embeddings, respectively,  are used to predict news veracity. Dedicated training objectives and data augmentation are utilised to optimise the model. \textbf{R} refers to the re-weighting algorithm which implicitly differentiates bias-aligned samples and bias-conflicting samples. Each color indicates a different data sample. }
\label{Fig.model}
\end{figure*}

In the section, we detail our model \method~for the cross-domain fake news detection task. \method\ is designed to extract and capitalise on subgraphs  from the news propagation graph. The architecture of \method~is illustrated in Fig.~\ref{Fig.model}. 

In \method, we take a small batch of propagation graphs and apply a mask generator on them to split each propagation graph into a casual subgraph and a biased subgraph. Then, the causal subgraphs and the biased subgraphs are encoded by two individual graph encoders, which produce two separate embeddings. The training objective is to emphasise the impact of the casual subgraphs while reducing the impact of the biased subgraphs on the fake news detection output.

For cross-domain detection, \method~is trained on  $D_{in}$ and then tested on $D_{out}$ in an unsupervised manner. When a few labelled samples are available from $D_{out}$, they can also be incorporated into the training process to further enhance the  model performance in the target domain. 


\subsection{Mask Generator}

Our mask generator learns a mask that splits each propagation graph $\mathcal{G}$ (i.e., $\mathcal{G}_k$ -- now we further drop the subscript `$k$' as long as the context is clear)
into a causal subgraph $\mathcal{G}_{c}$ and a biased subgraph $\mathcal{G}_b$. 
This is achieved by computing node importance scores (denoted as $\alpha_{i}$ for node $i$) and edge importance scores (denoted as  $\beta_{ij}$ for the edge between nodes $i$ and $j$) in the propagation graph $\mathcal{G}$ to measure the probability of a node or an edge belonging to the causal subgraph.

The mask generator takes graph $\mathcal{G}$ (i.e., its features) as input and outputs the importance of its nodes and edges. A Graph Isomorphism Network (GIN)~\cite{xu2018powerful} is utilised to encode the graph and map the node features $\mathbf{X}$ to node embeddings $\mathcal{H}$ for its superior graph structure representation ability. After obtaining the graph features $\mathcal{H} = \{{\mathbf{h}_{1}, \mathbf{h}_{2},\ldots,\mathbf{h}_{N}} \}$, where $N$ is the size of the node set and $\mathbf{h}_{i}$ represents the embedding for the $i$-th node, the node and edge importance scores are computed using an MLP: 
\begin{equation}
\label{Formula.node_masker}
\alpha_{i} =\sigma(\text{MLP}([\mathbf{h}_{i}])),         \beta_{ij} = \sigma(\text{MLP}([\mathbf{h}_{i}, \mathbf{h}_{j}]))
\end{equation}
Here, $\sigma$ is the activation function. 

Since the causal and the biased subgraphs are defined as two non-overlapping substructures of $\mathcal{G}$, the probability of a node and an edge belonging to  the biased subgraph can be established by $(1-\alpha_{i})$ and $(1-\beta_{ij})$, respectively.

Using the importance scores, we construct the causal graph mask $\mathbf{M}_{c} = [\mathbf{\alpha}, \mathbf{\beta}]$ and the biased graph mask $\mathbf{M}_{b} = [(1-\mathbf{\alpha}), (1-\mathbf{\beta})]$. Finally, the input propagation graph $\mathcal{G}$ is decomposed into a causal subgraph $\mathcal{G}_{c}=\{\mathbf{M}_{c} \odot \mathcal{G} \}$ and a biased subgraph $\mathcal{G}_{b}=\{\mathbf{M}_{b} \odot \mathcal{G}\}$, where $\odot$ is the filtering operation on the graph $G$ with the corresponding masks. The masks emphasize distinct regions of the propagation graphs, enabling subsequent GNN-based graph encoders to concentrate on different segments of the graphs.

\subsection{Graph Encoder}
\label{Sec.subgraph-encoder}

Two subgraph encoders, each of which is a two-layer Graph Convolutional Network (GCN)~\cite{kipf2016semi}, are used to encode the causal subgraph and the biased subgraph, respectively. 

Given a graph's node features $\mathbf{X} = \{\mathbf{x}_{1}, \mathbf{x}_{2},...,\mathbf{x}_{N}\}$ and its adjacency matrix $\mathbf{A}$, the graph embeddings are computed through GCNs by: 
\begin{equation}
\mathcal{Z}^{(l+1)} = \sigma\left(\mathbf{\tilde{D}}^{-1/2} \mathbf{\tilde{A}} \mathbf{\tilde{D}}^{-1/2} \mathcal{Z}^{(l)} \mathbf{W}^{(l)}\right)
\end{equation}
where $l=0$ or $1$, $\mathcal{Z}^{(0)}$ is the initial node features $\mathbf{X}$, $\mathbf{\tilde{A}}=\mathbf{A}+\mathbf{I}$ is the adjacent matrix of the graph with self-loops, $\mathbf{I}$ is the identity matrix, $\mathbf{\tilde{D}}$ is the degree matrix of $\mathbf{\tilde{A}}$, $\mathbf{W}^{(l)}$ is the learnable parameter matrix, and $\sigma$ is the activation function.

As shown in Figure~\ref{Fig.model}, two parallel subgraph encoders are used to encode the causal subgraph $\mathcal{G}_{c}$ and the biased subgraph $\mathcal{G}_{b}$ into a causal embedding $\mathbf{z}_{c}$ and a biased embedding $\mathbf{z}_{b}$. These embeddings will subsequently be fed into news classifiers for loss calculation and label (i.e., fake news or not) prediction.

\subsection{Classification Module}
The classification module (CM) is responsible for predicting the news veracity based on the extracted graph embeddings. It is composed of an MLP that uses the softmax function. Given the graph embedding $\mathcal{Z}$, which is the concatenation of $\mathbf{z}_{c}$ and $\mathbf{z}_{b}$, the CM acquires the prediction through: 
\begin{equation}
\label{eq.prediction-module}
pred = softmax(\text{MLP}(\mathcal{Z}))
\end{equation}

Since \method~focuses on classifying news according to causal features solely, we design a causal CM, denoted as $C_{c}$, and a biased CM, denoted as $C_{b}$, in the model. During model training, these two CMs are jointly trained to optimise \method\ to capture causal information accurately. In the test process, only the prediction results from the causal CM are used to detect fake news. More details about the use of the output of $C_{c}$ and $C_{b}$ are presented in the next subsection. 

\subsection{Disentangling Training Objectives}
\label{Sec.training-objectives}
In the training process, \method~is optimised in batches. As shown in Fig.~\ref{Fig.model}, a mini-batch of graphs are input into the model to calculate the loss for back propagation. 
Using the subgraph encoders, the casual subgraph $\mathcal{G}_{c}$'s embedding $\mathbf{z}_{c}$ and the biased subgraph $\mathcal{G}_{b}$'s embeddings $\mathbf{z}_{b}$ are computed.  The subgraph embeddings $\mathbf{z}_{c}$ and $\mathbf{z}_{b}$ are concatenated together as $\mathcal{Z} = \mathbf{z}_{c} \oplus \mathbf{z}_{b}$, to obtain the embeddings of the current batch of training data. To disentangle the two subgraphs $\mathcal{G}_{c}$ and $\mathcal{G}_{b}$, the following training objectives are defined. 

Note that, for both of the causal and biased subgraph extraction and classification, we optimise the model to predict final labels correctly, even though the biased part is not utilised in the testing process. The motivation is that the biased information is easily to be captured and also strongly correlated to the labels in the in-distribution data. The training of the mask generator requires the correct predictions of the biased classification module. 

\subsubsection{Loss for the Biased Branch}
Instead of traditional cross-entropy (CE) loss, the generalised cross-entropy (GCE) loss \cite{zhang2018generalized} is utilised to amplify the impact of the biased information and optimise the model to extract and embed the biased subgraph correctly: 
\begin{equation}
\label{Form.loss1}
\mathcal{L}_{biased}=\text{GCE}(C_{b}(\mathcal{Z}), y) = \frac{1-C_{b}^{y}(\mathcal{Z})^{q}}{q} 
\end{equation}
where $C_{b}(\mathcal{Z})$ and $C_{b}^{y}(\mathcal{Z})$ are the softmax outputs of the biased MLP classifier and the probability associated with it having the correct label $y$, respectively, and $q \in (0,1]$ here is a hyperparameter. 

Compared to the Cross-Entropy (CE) loss, the GCE loss can amplify the gradient of the standard CE loss for samples with a high confidence $C_{b}^{y}(\mathcal{Z})$ of predicting the correct label $y$~\cite{zhang2018generalized}. As a result, the GCE loss is able to amplify the impact of biased information and optimise the biased branch to extract and embed the biased subgraph correctly. 

\subsubsection{Loss for the Causal Branch}
For the causal subgraph encoder, we train it with the MLP classifier $C_{c}$ by a weighted CE loss based on: 
\begin{equation}
\label{Form.loss2}
\mathcal{L}_{causal}=\text{CE}(C_{c}(\mathcal{Z}))_{weighted} = W(\mathcal{Z})\cdot \text{CE}(C_{c}(\mathcal{Z}))
\end{equation} 
where CE is the standard cross-entropy loss, and the weight is defined as: 
\begin{equation}
\label{Form.loss2-weight}
W(\mathcal{Z}) = \frac{\text{CE}(C_{b}(\mathcal{Z}), y)}{\text{CE}(C_{b}(\mathcal{Z}), y) + \text{CE}(C_{c}(\mathcal{Z}), y)}
\end{equation}
This weight is based on the fact that graphs with a high CE loss (worse prediction) from the biased MLP classifier $C_{b}$ can be considered as containing more causal (instead of biased) information. 

Overall, for the disentanglement of the causal and the biased substructures, the loss $\mathcal{L}_{dis}$ is the sum of the losses  $\mathcal{L}_{biased}$ and $\mathcal{L}_{causal}$, from the causal and biased Classification Modules, respectively.
\begin{equation}
\label{equation.L-dis}
\mathcal{L}_{dis} =  \mathcal{L}_{biased} + \mathcal{L}_{causal}
\end{equation}
Next, we consider the connection between the causal and the biased subgraphs. 

\subsubsection{Batch-wise Data Augmentation}

The causal and biased subgraphs of the same graph are inherently correlated. To learn and mitigate the correlations between the casual and the biased embeddings $\textbf{z}_{c}$ and $\textbf{z}_{b}$,  inspired by  a previous work~\cite{lee2021learning}, we use a data augmentation strategy that  randomly permutes the bias vectors in a batch to obtain $\mathcal{Z}_{unbiased} = \textbf{z}_{c} \oplus (\hat{\textbf{z}}_{b})$, where  $(\hat{\textbf{z}}_{b})$ are the permuted bias vectors. The augmented embeddings $\mathcal{Z}_{unbiased}$ have less correlations between the causal embeddings $\textbf{z}_{c}$ and the biased embeddings $\textbf{z}_{b}$ because of the vector swapping, since now each news' causal embedding is concatenated with another news' biased embedding.

Similarly, $\mathcal{Z}_{unbiased}$ is fed into the CMs $C_{c}$ and $C_{b}$ to predict the news veracity labels. For the causal part $C_{c}(\mathcal{Z}_{unbiased})$, the loss calculation follows the previous weighted CE loss in Formula (\ref{Form.loss2}) and (\ref{Form.loss2-weight}): 
\begin{equation}
\label{Form.loss3}
\begin{aligned}
\mathcal{L}_{causal}^{aug}=W(Z_{unbiased})\cdot \text{CE}(C_{c}(\mathcal{Z}_{unbiased}))    
\end{aligned}
\end{equation}

For the biased part $C_{b}(\mathcal{Z}_{unbiased})$, to ensure the causal and biased substructures the biased subgraph encoder and $C_{b}$ still focus on the biased information, we also permute the labels $y$ in the same order as $\hat{\textbf{z}}_{b}$ above to generate $\hat{y}$. The GCE loss then becomes:
\begin{equation}
\label{From.loss4}
\mathcal{L}_{biased}^{aug} = \text{GCE}(C_{b}(\mathcal{Z}_{unbiased}), \hat{y})
\end{equation}

For the decorrelation of causal and biased embeddings, the loss $\mathcal{L}_{swap}$ is defined as: 
\begin{equation}
\label{Form.loss-swap}
\begin{aligned}
\mathcal{L}_{swap} = \mathcal{L}_{causal}^{aug} + \mathcal{L}_{biased}^{aug}
\end{aligned}
\end{equation}

Finally, the overall loss function $\mathcal{L}$ for  training \method\ is given as the sum of the disentanglement loss and the decorrelation loss from Formula (\ref{equation.L-dis}) and (\ref{Form.loss-swap}): 
\begin{equation}
\label{Eq.Ori-loss}
\mathcal{L} = \mathcal{L}_{dis} + \mathcal{L}_{swap}
\end{equation}

During training, to prevent gradients from the causal branch from influencing the biased graph encoder, only gradients from $z_{c}$ are allowed to flow back to the causal classification module, $\mathcal{Z}$ and $\mathcal{Z}{unbiased}$. Gradients associated with $z{b}$ and $\hat{z_{b}}$ are detached to block them from propagating. Similarly, gradients from the biased branch are kept separate from the causal components.

\subsection{Model Fine-tuning with OOD Data}\label{subsec:enhancement}

\method\ can be trained using just in-distribution data $D_{in}$. We can also use a few labelled OOD data samples for further fine-tuning \method\ on a target domain. In this subsection, we discuss model optimisation given the availability of a few OOD samples via a contrastive learning method. 

When limited OOD data samples are available, we can further improve the model performance via aligning the representation space of causal fake news. This is achieved by making the representations of in-distribution and OOD samples from the same veracity class closer while keeping representations from different classes further away. We achieve this via contrastive learning.

The in-distribution data samples are the primary resource used for model training, while the OOD data samples occur much less frequently. We therefore need to learn more separated news representations for the in-distribution data from different classes (i.e. true or fake news). 
To achieve this goal, we adopt a supervised contrastive learning objective to bring closer samples from the same class and separate different classes among the in-distribution samples. This is given as: 
\begin{equation}
\label{Equation.CL1}
\begin{aligned}
\mathcal{L}_{CL}^{in} = -\frac{1}{N^{in}} \sum_{n=1}^{N^{in}}\frac{1}{N_{y_{n}^{in}}}\sum_{m=1}^{N^{in}}\mathds{1}_{[n\ne m]}\mathds{1}_{[y_{n}^{in}= y_{m}^{in}]} \\ \log \frac{\exp({sim(o_{n}^{in}, o_{m}^{in})/\tau})}{\sum_{k=1}^{N^{in}}\mathds{1}_{[n\ne k]} \exp(sim(o_{n}^{in}, o_{k}^{in})/\tau) }
\end{aligned}
\end{equation}
where $N^{in}$ is the number of in-distribution data samples in a batch, $N_{y_{n}^{in}}$ is the number of in-distribution data samples which share the same label $y_{n}^{in}$ with sample $C_{n}^{in}$, $\mathds{1}$ is the indicator function, $o_{n}^{in}$, $o_{m}^{in}$, and $o_{k}^{in}$ are the corresponding extracted casual representations from \method, $sim(\cdot)$ is the cosine similarity function, and $\tau$ is a hyperparameter that controls the temperature. 

To fine-tune \method\ over OOD data, another supervised contrastive learning objective is proposed. Here, we aim to draw the embedding space of samples with the same label but from different distributions closer. 
\begin{equation}
\label{Equation.CL2}
\begin{aligned}
\mathcal{L}_{CL}^{out} = -\frac{1}{N^{out}} \sum_{n=1}^{N^{out}}\frac{1}{N_{y_{n}^{out}}}\sum_{m=1}^{N^{in}}\mathds{1}_{[y_{n}^{out} = y_{m}^{in}]} \\ \log \frac{\exp({sim(o_{n}^{out}, o_{m}^{in})/\tau})}{\sum_{k=1}^{N^{in}} \exp(sim(o_{n}^{out}, o_{k}^{in})/\tau) }
\end{aligned}
\end{equation}
where $N^{out}$ is the number of OOD samples in a training batch, $N^{in}$ is the number of in-distribution samples in the batch, $N_{y_{n}^{out}}$ is the number of in-distribution samples which share the same label $y_{n}^{out}$ with sample $C_{n}^{out}$, and $o_{n}^{out}$, $o_{m}^{in}$, and $o_{k}^{in}$ are the corresponding extracted causal representations from \method.

Loss $\mathcal{L}_{CL}$ for the contrastive learning is then given as the sum of Equations~(\ref{Equation.CL1}) and (\ref{Equation.CL2}): 
\begin{equation}
\mathcal{L}_{CL} = \mathcal{L}_{CL}^{in} + \mathcal{L}_{CL}^{out}
\end{equation}

In summary, the overall training objective $\mathcal{L}_{en}$ for \method~is a weighted sum of the contrastive learning loss and the original loss $\mathcal{L}$ as shown in Equation~(\ref{Eq.Ori-loss}).
\begin{equation}
\mathcal{L}_{en} = \gamma \cdot \mathcal{L} + (1-\gamma) \cdot \mathcal{L}_{CL}
\end{equation}
Here, $\gamma$ is a hyperparameter controlling the contribution of the contrastive learning loss. 

\section{Experiment}

\subsection{Experimental Settings}
\label{Sec.Experiment-Settings}
\textbf{Datasets.} 
Four public datasets collected from Twitter (now called X) and Weibo (a Chinese social media platform like Twitter) are utilised in the experiment: (1)~\texttt{Twitter}~\cite{ma2017detect}, (2)~\texttt{Weibo}~\cite{ma2016detecting}, (3)~\texttt{Twitter-COVID19}~\cite{lin-naacl-2022-aclr} and (4)~\texttt{Weibo-COVID19}~\cite{lin-naacl-2022-aclr}. The statistics of the datasets are shown in Table~\ref{Tab.data-stats}. 

 \texttt{Twitter} and \texttt{Weibo} are open-domain datasets. They cover a variety of topics except COVID-19 and are used as the main training set. \texttt{Twitter-COVID19} and \texttt{Weibo-COVID19} only contain news related to COVID-19, which represent the OOD data in the experiments. 
 To support the required domain adaptation, a subset of COVID19 data samples are also selected for the fine-tuning purposes in the second set of experiments.

To showcase the effectiveness of the proposed model \method, two set of experiments are designed.

In the first set of experiments, the models are trained on in-distribution data (e.g., \texttt{Twitter}) and tested on OOD data (e.g., \texttt{Twitter-COVID19}), to  simulate the scenario where no prior knowledge about the OOD data is available.

In the second set of experiments, a few OOD samples (e.g., 20\% of \texttt{Twitter-COVID19}) are utilised to help optimise the models together with in-distribution data (e.g., \texttt{Twitter}), to simulate the scenario where we have a small number of manually labelled OOD samples, which could happen after an explosion of some hot news. The remaining OOD data (e.g, 80\% of \texttt{Twitter-COVID19}) are used for model testing.


\begin{table}[t]
\small
\setlength\tabcolsep{1.0pt}
\centering
\caption{Experimental Dataset Statistics  (``Avg. depth'' refers to the average number of layers of the news propagation graphs, i.e., trees)}
\label{tab:statistics}
\begin{tabular}{@{}lcccc@{}}
\toprule
& \textbf{Twitter} & \textbf{Twitter-COVID} & \textbf{Weibo} & \textbf{Weibo-COVID} \\
\midrule
\# news           & 1,154    & 400    & 4,649   & 399    \\
\# graph nodes     & 60,409   & 406,185  & 1,956,449 & 26,687\\
\# true news       & 579     & 148    & 2,336   & 146 \\
\# fake news           & 575     & 252    & 2,313   & 253    \\
Avg. depth        & 11.67   & 143.03   & 49.85  & 4.31\\
Avg. \# posts  & 52      & 1,015     & 420    & 67 \\
Domain                 & Open    & COVID-19 & Open   & COVID-19 \\
Language               & English & English & Chinese & Chinese \\
\bottomrule
\end{tabular}
\label{Tab.data-stats}
\end{table}

\noindent\textbf{Baselines.} We compare with 11 models including two SOTA models CADA~\cite{li-acmmm-2023improving} and ACLR~\cite{lin-naacl-2022-aclr}.

Baseline models trained with in-distribution data only: 
\textbf{LSTM} \cite{ma2016detecting} uses an LSTM-based model to learn feature representations of relevant posts over time. \textbf{CNN} \cite{yu2017convolutional} uses a CNN model for misinformation identification by modelling the relevant posts as a fixed-length sequence. 
\textbf{RvNN} \cite{ma2018rumor} learns the propagation of news by exploiting a tree structured recursive neural network. \textbf{PLAN} \cite{khoo2020interpretable} uses a Transformer~\cite{vaswani2017attention}-based model for fake news detection by capturing long-distance interactions between tweets (source post and comments). 
\textbf{RoBERTa}~\cite{liu2019roberta} encodes the text information of a news item  and classifies the news with text classification. 
\textbf{BiGCN}~\cite{bian2020rumor} models news propagation by representing social media posts as nodes in a graph. It then utilises a GCN-based model to encode the graph and classifies if a given news item is true or fake.
\textbf{GACL}~\cite{sun2022rumor} enhances BiGCN~\cite{bian2020rumor} by generating adversarial training samples and training based on contrastive learning. 
\textbf{SEAGEN}~\cite{gong2023fake} models the  news propagation process by encoding the temporal propagation graph with a temporal graph network (TGN) and a neural Hawkes process, which is used for fake news detection.
\textbf{UCD-RD} \cite{ran-aaai-2023-unsupervised} uses prototype-based contrastive learning to initialise prototypes via in-distribution samples, and aligns the OOD data features with the corresponding prototypes.

Baseline models trained with both in-distribution and low-resource OOD data: \textbf{ACLR}~\cite{lin-naacl-2022-aclr} utilises adversarial contrastive learning to transfer pre-trained BiGCN~\cite{bian2020rumor} models from a source domain to a target domain for fake news detection. 
\textbf{CADA}~\cite{li-acmmm-2023improving} serves as a plugging-in module and adapts pre-trained models from a source domains to a target domain by label-aware domain adversarial neural networks~\cite{ganin-icml-2015-dann-unsupervised}. In our experiment, it is combined with BiGCN, RoBERTa, SEAGEN and GACL as the pre-trained models.

\noindent\textbf{Implementation and Parameter Settings}
All baselines and our model \method~are implemented in Pytorch\footnote{https://pytorch.org/} and trained on GPU A100. The baseline models use the default hyperparameter settings from their original papers. Hyperparameter $\gamma$, $q$, $\tau$ of our \method\ model is set to 0.2, 0.7, 0.1 respectively in the experiments to present the final results. The hyperparameters are selected empirically with grid search. 

\begin{table}[t]
    \caption{Fake News Detection Methods' Few-Shot Performance on \texttt{Twitter-COVID19} and \texttt{Weibo-COVID19} (Acc: Accuracy score on fake news detection; T:True news; F:Fake news)}
    \centering
    \setlength{\tabcolsep}{1mm}{
    \begin{tabular}{l | c c c|   c c c }
        \toprule
        Method & \multicolumn{3}{c|}{\texttt{Twi}$\rightarrow$\texttt{Twi-COVID}}& \multicolumn{3}{c}{\texttt{Wei}$\rightarrow$\texttt{Wei-COVID}} \\
        \cline{2-7}& Acc &  T-F1  & F-F1 & Acc & T-F1  & F-F1 \\\midrule
        $\text{CADA}_\text{BiGCN}$ & 0.681 & 0.621 & 0.725  & 0.716 & 0.552 & 0.792\\
        $\text{CADA}_\text{RoBERTa}$ & 0.711 & 0.540 & 0.790 & 0.839 & 0.783 & 0.878 \\ 
        $\text{CADA}_\text{SEAGEN}$ &  0.669 & 0.383 & 0.785 & 0.662 & 0.471 & 0.752 \\
        $\text{CADA}_\text{GACL}$  & 0.641 & 0.511 & 0.716 &  0.684 & 0.402 & 0.786  \\
        ACLR & \underline{0.741} & \underline{0.607} & \textbf{0.799} & \underline{0.897} & \underline{0.847} & \underline{0.917} \\ 
        $\text{\method}_\text{Fine-Tuned}$ & \textbf{0.772} & \textbf{0.767} & \underline{0.797} & \textbf{0.922} & \textbf{0.884} & \textbf{0.940} \\  \hline
        $\uparrow$ (\%) & +4.18 & +26.36 & -0.25 & +2.79 & +4.37 & +2.51 \\        
        \bottomrule
    \end{tabular}}
    \label{Tab.small-res}
\end{table}

\begin{table*}[t]
    \caption{Fake News Detection Methods' Zero-Shot Performance on \texttt{Twitter-COVID19} and \texttt{Weibo-COVID19} (Acc: Accuracy score on fake news detection; F-F1: F1 score on fake news detection; T-F1: F1 score on true news detection)}
    \centering
    \setlength{\tabcolsep}{1.7mm}{
    \begin{tabular}{l | c c c| c c c | c c c| c c c }
        \toprule
        Source & \multicolumn{6}{c|}{\texttt{Twitter}}& \multicolumn{6}{c}{\texttt{Weibo}} \\ \midrule
        Target & \multicolumn{3}{c|}{\texttt{Twitter-COVID19}} & \multicolumn{3}{c|}{\texttt{Weibo-COVID19}} & \multicolumn{3}{c|}{\texttt{Twitter-COVID19}} & \multicolumn{3}{c}{\texttt{Weibo-COVID19}}\\ \midrule
        Method & Acc &  T-F1  & F-F1 & Acc & T-F1  & F-F1 & Acc &  T-F1  & F-F1 & Acc & T-F1  & F-F1 \\\midrule

        LSTM   & 0.412 & 0.426 & 0.340 &0.463 & 0.329 & 0.498 & 0.510 & 0.243 & 0.533 &0.416 & 0.428 & 0.416 \\
        CNN    & 0.406 & 0.450 & 0.285 & 0.445 & 0.328 & 0.476 & 0.498 & 0.249 & 0.528 & 0.421 & 0.438 & 0.382 \\
        RvNN& 0.436 & 0.458 & 0.401 & 0.514 & 0.426 & 0.538 & 0.540 & 0.247 & 0.534 & 0.479 & 0.548 & 0.437 \\
        PLAN   & 0.455 & 0.432 & 0.476 & 0.532 & 0.414 & 0.578 & 0.573 & 0.298 &0.549 & 0.384 & 0.283 & 0.461 \\
        RoBERTa& 0.479 & 0.430 & 0.531 &  0.623 & 0.459 & \underline{0.711} & 0.603 & \textbf{0.585} & 0.619 & 0.680 & 0.714 & 0.637 \\ \midrule
        BiGCN  & 0.468 & \underline{0.546} & 0.356 & 0.569 & 0.429 & 0.586 & 0.616 & 0.252 & 0.577 & 0.612 & \underline{0.681} & 0.441 \\
        SEAGEN & 0.494 & 0.448 & 0.494 &  0.555 & 0.406 & 0.583 & 0.578 & 0.320 & 0.650 &  0.586  & 0.613 & 0.424 \\
        GACL   & 0.541 & 0.545 & 0.536 &   0.601 & 0.410 & 0.616 &  \underline{0.621} & 0.345 & \underline{0.666} & 0.688 & 0.635 & 0.727 \\ 
        UCD-RD   & \underline{0.665} & 0.453 & \underline{0.762} & \underline{0.631} & \underline{0.510} & 0.621 & 0.591 & 0.371 & 0.583 & \underline{0.689} & 0.451 & \underline{0.783} \\ 
        \method~(ours)& \textbf{0.725} & \textbf{0.583} & \textbf{0.782} &  \textbf{0.732} & \textbf{0.608} & \textbf{0.796} & \textbf{0.672} & \underline{0.563} & \textbf{0.741} & \textbf{0.742} & \textbf{0.721} & \textbf{0.809} \\
        \hline
        $\uparrow$ (\%) & +9.02 & +6.78 & +2.62 & +16.00 & +19.22 & +11.95 & +8.21 & -3.76 & +11.26 & +7.69 & +5.87 & +3.32 \\ \bottomrule

    \end{tabular}}
    \label{Tab.res1}
\end{table*}
\vspace{-2mm}
\subsection{Results}
Table~\ref{Tab.small-res} and Table~\ref{Tab.res1} present the model performance on the four dataset settings (from \texttt{Twitter} to \texttt{Twitter-COVID19}, \texttt{Weibo-COVID19} and from \texttt{Weibo} to \texttt{Twitter-COVID19}, \texttt{Weibo-COVID19}). 

In Table~\ref{Tab.res1}, the models are divided into two groups. Models in the upper group are sequence-based models (LSTM, CNN, RvNN, PLAN and RoBERTa) while models in the bottom group are graph-based models (BiGCN, SEAGEN, GACL, UCD-RD and \method). The graph-based models generally perform better than the sequence-based ones, which shows the effectiveness in utilising propagation graphs for fake news detection. Among the graph-based models, our \method\ model performs the best consistently over both datasets in terms of both accuracy and most F1 scores. The baseline models without considering OOD data generally perform poorly. These models are trained on the open-domain in-distribution datasets and have been biased by domain-specific information. UCD-RD attempts to align the in-distribution news representations and the OOD news representations for samples of the same class. It does not consider the casual substructures and hence is still outperformed by \method, with a performance gap of $7\%\sim16\%$.

As shown in Table.~\ref{Tab.small-res}, when labelled OOD  data is available, the baseline models (BiGCN, RoBERTa, SEAGEN and GACL) powered by CADA can learn features from the OOD data and achieve better accuracy than their vanilla version. ACLR which is designed  with domain adaptation in mind achieves even better performance. However, these models are still outperformed by 
\method\ using fine-tuning in most cases, with a performance gap around $2.79\%\sim 4.18\%$.

\subsection{Ablation Experiment}
To show the effectiveness of the causal subgraph extraction module and each loss function, five variants of \method~are trained and the performance is presented in Table~\ref{Tab.ablation-res}. In the first variant ``\textbf{No-Causal}'', the casual subgraph extraction module is removed, and we train the model purely with the cross entropy loss. The remaining four variants all use causal subgraph extraction. They each add one more loss component, with the final model being the complete \method\ model. The two columns represent experiments with \texttt{Twitter->Twiter-COVID19}  and \texttt{Weibo->Weibo-COVID19} data. The results show the important of each model components especially the causal subgraph extraction module, which enables the additional loss functions that together yield  the substantial improvements achieved by \method\ over the SOTA models. 
\vspace{-2mm}
\begin{table}[ht]
\small
\setlength\tabcolsep{5.0pt}
\centering
\caption{Ablation Experiment Results for Our \method\ Model}
\label{Tab.ablation-res}
\begin{tabular}{@{}lcc@{}}
\toprule
\textbf{Model variants} & \texttt{Twitter} & \texttt{Weibo}\\ 
\midrule
No-Causal & 0.468 & 0.612 \\
+$\mathcal{L}_{biased}$ & 0.502  & 0.647 \\
+$\mathcal{L}_{causal}$ & 0.656 & 0.698 \\
+$\mathcal{L}_{causal}^{aug}$ & 0.688 & 0.726 \\
+$\mathcal{L}_{biased}^{aug}$
 & \textbf{0.725} & \textbf{0.742} \\
\bottomrule
\end{tabular}
\end{table}
\vspace{-3mm}
\subsection{Case Study}


The effectiveness of \method~is demonstrated through a case study using the Twitter and Twitter-COVID19 datasets. The mask generator, trained on the \texttt{Twitter} dataset, is applied to the \texttt{Twitter-COVID19} dataset to filter out biased subgraphs while preserving causal ones. As shown in Fig.\ref{Tab.Case-Study}, the source news, which mimics an official tone, is difficult for linguistic-based models to classify and receives a low node score, indicating its content alone is insufficient for accurate classification. In contrast, comments that offer insights into the news veracity are assigned higher node/edge scores, while unrelated content like propaganda is scored lower. This differentiation allows \method's Graph Encoder to focus on causal information, thereby improving detection performance.

\vspace{-2mm}
\begin{table}[ht]
\small
\setlength\tabcolsep{5.0pt}
\centering
\caption{Case Study Example from \texttt{Twitter-COVID19} dataset. (The indexes of news/comments are specified by the index number. The node scores and edge scores are calculated by \method's mask generator)}
\begin{tabular}{p{8cm}}
\toprule
\textbf{News, Comments, Node Scores and Edge Scores} \\ 
\midrule
\textbf{News 0}: The World Health Organization confirmed that Covid-19 is deadlier than the seasonal flu, but does not transmit as ... [Node Score: \textless 0.001] \\ \midrule
\textbf{Comment 1}: Need to buy a lot of masks contact me. [Node Score: 0.154] [Edge 0$\rightarrow$1 Score: 0.152] \\\midrule
\textbf{Comment 2}: Because of their more rigorous testing protocols, South Korea’s mortality rate of 0.6\% is the most accurate. [Node Score: 0.393] [Edge 1$\rightarrow$2 Score: 0.515] \\ \midrule
\textbf{Comment 3}: why don't you look at implementing \#Covid\_19 travel health cards that confirm the person has been… [Node Score: 0.514] [Edge 0$\rightarrow$3 Score: 0.462] \\ \midrule
\textbf{Comment 4}: WHO is also omitting mild cases from their stats. [Node Score: 0.556] [Edge 1$\rightarrow$4 Score: 0.574] \\ \midrule 
More comments and conversations... \\
\bottomrule
\end{tabular}
\label{Tab.Case-Study}
\end{table}

\vspace{-4mm}
\section{Conclusion}
We proposed a model named \method\ for detecting fake news across domains by extracting and leveraging causal substructures. \method~addresses the limitations of existing models in handling domain biases and OOD data, highlighting the importance of causal elements in news propagation graphs. Through extensive experiments, we show that \texttt{CSDA} outperforms not only  sequence-based models but also other graph-based models, achieving higher accuracy, particularly in cross-domain scenarios. Additionally, the integration of a fine-tuning process with low-resource OOD data further enhances \method's robustness and adaptability.

For future work, it would interesting to further exploit  the causal information from the textual content of the news. 
\bibliography{aaai25}

\end{document}